\documentclass[a4paper,10pt]{article}

\usepackage[affil-it]{authblk}  

\usepackage{graphicx}  
\usepackage{amsmath}   
\usepackage{amssymb}   

\usepackage{color}

\usepackage[active]{srcltx}

\def\eq#1{{Eq.~(\ref{#1})}}

\newcommand{\LL}{Lanczos-Lovelock }

\title{Cosmic Information,  the Cosmological Constant and the Amplitude of primordial perturbations}

\author[1]{T. Padmanabhan\thanks{paddy@iucaa.in}}
\affil{\small IUCAA, Pune University Campus,
  Ganeshkhind, Pune - 411 007, India.}

\author[2]{Hamsa Padmanabhan\thanks{hamsa.padmanabhan@phys.ethz.ch}}
\affil{\small Institute for Astronomy, ETH Zurich, Wolfgang-Pauli-Strasse 27, CH-8093 Zurich, Switzerland.}

\date{ }

\begin{document}

\maketitle

\begin{abstract}
A unique feature of gravity is its ability to control the information accessible to any specific observer. We quantify the notion of  cosmic information (`CosmIn') for an eternal observer in the universe. Demanding the finiteness of CosmIn \textit{requires} the universe to have a late-time accelerated expansion.  Combining the introduction of CosmIn with generic features of the quantum structure of spacetime (e.g., the holographic principle), we present a holistic model for cosmology. We show that (i) the numerical value of the cosmological constant, as well as (ii) the amplitude of the primordial, scale invariant, perturbation spectrum can be determined in terms of a single free parameter, which specifies the energy scale at which the universe makes a transition from a pre-geometric phase to the classical phase. For a specific value of the parameter, we obtain the correct results for both (i) and (ii). This formalism also shows that the quantum gravitational information content of  spacetime can be  tested using precision cosmology.
\end{abstract}

\maketitle

It is now well established that information is a physical entity \cite{landauer} and the flow of information has concrete physical consequences.
The fact that gravity controls the amount of spacetime information accessible to a given observer, suggests that one can acquire deeper insights into spacetime dynamics through its information content. The concept of information, being a common ingredient in both classical and quantum regimes,  can thus be used to provide a link between the  descriptions of spacetime in these two domains.  

The key difficulty in formulating this connection lies in quantifying the amount of spacetime information. While this is indeed difficult for a \textit{general} spacetime, we show that it is possible to introduce a natural definition of information content  in the context of \textit{cosmological} spacetimes (`CosmIn') and use it to link the quantum and classical phases of the universe. Moreover, we shall see that this information paradigm allows us to determine both, (i) the numerical value of the cosmological constant and (ii) the amplitude of the primordial, scale invariant, power spectrum of  perturbations, thus providing a holistic descrption of cosmology.

In any Friedmann model, the \textit{proper} length-scales (say, the wavelengths of the modes of a field) scale as $\lambda(a) \propto a$ and can cross the \textit{proper} Hubble radius $H^{-1}(a)=(\dot a/a)^{-1}$ as the universe evolves. 
The number of modes  $dN$  located in the comoving Hubble volume $V_H(a) = (4\pi/3) (aH)^{-3}$, which have comoving wave numbers in the range  $d^3k$, is given by $dN = V_H(a) d^3k/(2\pi)^3\equiv V_H(a) dV_k/(2\pi)^3$ where $dV_k=4\pi k^2 dk$. 
A mode with a comoving wave number $k$ crosses the Hubble radius when $k=k(a)\equiv a H (a)$. So, the modes with wave numbers between  $k$ and $k+dk$, where $dk= [d(aH)/da]\, da$, cross the Hubble radius 
during the interval ($a, a+da$). We \textit{define} the information associated with  modes which cross the Hubble radius during any interval $a_1<a<a_2$ by 
\begin{equation}
N(a_2,a_1) = \pm\int_{a_1}^{a_2} \frac{V_H(a)}{(2\pi)^3} \, \frac{dV_k[k(a)]}{da} \, da = \pm\frac{2}{3\pi}\ln \left( \frac{h_1}{h_2}\right)
\label{defN}
\end{equation} 
where $h(a)\equiv H^{-1}(a)/a$ is the \textit{comoving} Hubble radius and $h_1=h(a_1),h_2=h(a_2)$. The sign is chosen to keep $N$ positive, by definition. 

In the absence of any untested physics from the matter sector (like e.g., inflationary scalar fields, which we will \textit{not} invoke in this paper), the universe  is radiation dominated at early epochs and, classically,  has a singularity at $a=0$.  In reality, the classical description breaks down when quantum gravitational effects set in. 
We assume that the universe makes a transition from a quantum, pre-geometric phase to the classical, geometric phase at an epoch $a=a_{\rm QG}$ when the radiation energy density  is $\rho_R=\rho_{\rm QG}$ where $(8\pi/3)\rho_{\rm QG} \equiv  E_{\rm QG}^4$. We express  the energy scale as $E_{\rm QG}\equiv\nu^{-1} E_{\rm Pl}$ 
where $E_{\rm Pl}\equiv \hbar c/L_{\rm P}=1/L_{\rm P}$ in natural units  ($\hbar=1=c$) and $L_P\equiv(G\hbar/c^3)^{1/2}=G^{1/2}$ is the Planck length; $\nu$ is a numerical factor which, as we shall see, can be determined from observations \cite{comment1}. The Hubble radius at $a=a_{\rm QG}$ is $H_{\rm QG}^{-1}\equiv \nu^2 L_P$.  

If the universe was populated by sources which satisfy $(\rho+3p)>0$ for all $a>a_{\rm QG}$, then the function $N(a,a_{\rm QG})$, defined by \eq{defN}, is a monotonically increasing function of $a$ and \textit{diverges} as $a\to\infty$. 
It is reasonable to demand that \textit{$N(a,a_{\rm QG})$ should be finite and its finite value should be determined by purely quantum gravitational considerations.} This would require the comoving Hubble radius $H^{-1}(a)$ to reach a maximum value  at some epoch, say, $a=a_\Lambda$. Then  the number of modes $N(a_\Lambda,a_{\rm QG})$ which \textit{enter} the Hubble radius during the entire history of the universe --- which we call `CosmIn' --- will be a finite constant, say $N(a_\Lambda,a_{\rm QG})\equiv I_c$. This, in turn, requires $\rho+3p=0$ at $a=a_\Lambda$ with $\rho+3p<0$ for $a>a_\Lambda$. \textit{The finiteness of CosmIn thus demands that we must have an accelerating phase in the universe.}

This finiteness of CosmIn is closely related to the finiteness of another observable, $x(a_2,a_1)$ which is the maximum \textit{comoving} distance a signal can propagate during the time interval $a_1<a<a_2$. An eternal observer (that is, an observer located at the origin and making observations at  very late times) will be able to receive signals emitted at epoch $a$ from a maximum comoving distance 
\begin{equation}
 x(\infty,a)\equiv x_\infty(a) = \int_{t}^{\infty} \frac{dt}{a(t)} = \int_a^\infty \frac{d\bar a}{\bar{a}^2 H(\bar a)}
\label{xinfin}
\end{equation} 
 In particular, the maximum comoving distance the eternal observer can probe on the spatial hypersurface $a=a_{\rm QG}$ --- which corresponds to the birth of the classical spacetime --- is given by $x_\infty(a_{\rm QG})$. If $x_\infty(a_{\rm QG})$ is divergent, then such an observer can access information from an infinite region of space at $a=a_{\rm QG}$. On the other hand, if $x(\infty,a_{\rm QG})$ is finite, then  the size of the cosmic space which the eternal  observer can access on the surface  $a=a_{\rm QG}$ will be finite, and there is an \textit{information horizon}. 
From \eq{xinfin}, it is easy to see that if the universe was populated by sources which satisfy $(\rho+3p)>0$ for all $a>a_{\rm QG}$, then $x_\infty(a_{\rm QG})$ is divergent. In fact, as long as $(\rho+3p)>0$ \textit{asymptotically} (i.e., as $a\to\infty$), then $x(\infty, a)$ is divergent for \textit{all} $a$. On the other hand, an accelerated phase, due to $(\rho+3p)<0$ for all $a>a_\Lambda$
will ensure that $x_\infty(a_{\rm QG})$ is also finite. 

 The simplest way to ensure that $(\rho+3p)<0$ at late times without invoking untested physics (like e.g., quintessence) is to introduce a non-zero cosmological constant, with energy density $\rho_\Lambda$.  The expansion of such a universe, for $a>a_{\rm QG}$,  is driven by the energy density of matter $\rho_m \propto a^{-3}$, radiation $\rho_R \propto a^{-4}$ and the cosmological constant $\rho_\Lambda$. Defining the density $\rho_{\rm eq} \equiv \rho_m^4(a)/\rho_R^3(a)$ which is a constant independent of $a$, we can model the universe as a dynamical system described by three densities: ($\rho_{\rm QG},\rho_{\rm eq},\rho_\Lambda$). 

\begin{figure}[t]
\scalebox{0.31}{\input{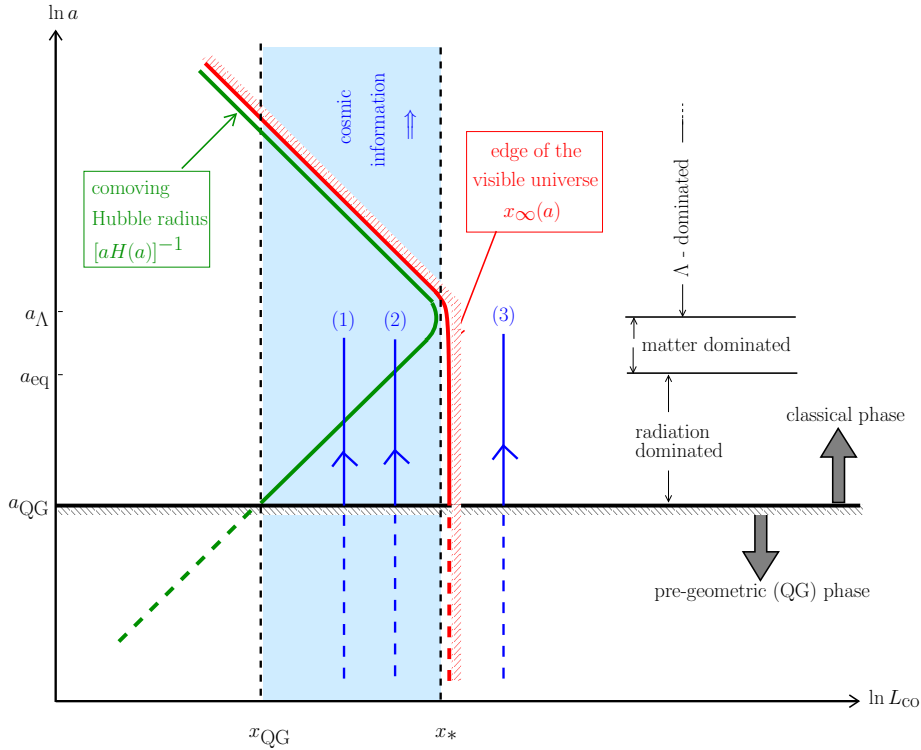}}
 \caption{Various length scales of interest in cosmological evolution. See text for discussion.}
\label{fig:lengthscales}
\end{figure}

Figure~\ref{fig:lengthscales} shows the behaviour  of  the comoving Hubble radius $h(a)\equiv H^{-1}(a)/a$ (green line) and $x_\infty(a)$ (red line) schematically (i.e., not to scale) for such a universe. The comoving Hubble radius increases during the radiation dominated ($h\propto a$) and matter dominated ($h\propto a^{1/2}$) phases and decreases ($h\propto a^{-1}$) in the cosmological constant dominated phase. The turn-around occurs at the epoch $a=a_\Lambda$. The classical description loses its relevance  at $a=a_{\rm QG}$; this limit is shown as a horizontal (black) line at $a=a_{\rm QG}$. 

Somewhat surprisingly, the functional form of $x_\infty(a)$ has not attracted the attention it deserves. We see that during the phase dominated by the cosmological constant, $x_\infty(a)$ decreases as $1/a$. But at earlier times, 
$x_\infty(a)$
remains very nearly constant (changing only by a factor 3    
when  $a$ changes by nearly a factor $3000$). 
The signals travel a \textit{finite} comoving distance $x_*\equiv x_\infty(0)$ during the \textit{entire history}, $0<t<\infty$  of the universe \cite{comment3}. 

Observations indicate that  $\rho_{\rm eq} = [0.86 \pm 0.09 \  {\rm{eV}}]^4$ and $\rho_\Lambda  = [(2.26 \pm 0.05) \times 10^{-3}  \ {\rm{eV}}]^4$. The theoretical status of these numerical values of $\rho_{\rm eq}$ and $\rho_\Lambda$ are very different. The  value of $\rho_{\rm eq}$ depends on the nature and abundance of dark matter and baryons relative to  photons and --- in principle --- can be determined from high-energy physics.  But, as is well-known, we do not have any theoretical basis to  determine  $\rho_\Lambda$ which  is considered a major challenge in theoretical physics. 

However, in our approach, the value of $\rho_\Lambda$ is determined by the value of  $N(a_\Lambda,a_{\rm QG})\equiv I_c$. 
The calculation of $I_c$ is completely straightforward but a bit tedious.  (See Appendix C of \cite{hptpreview} for details.) The final result is given by:
\begin{equation}
 I_c = -\frac{2}{3\pi}\ln\left[\frac{k_1 (\rho_\Lambda^2\rho_{\rm eq})^{1/12}}{E_{\rm QG}} \right]
 =\frac{2}{3\pi}\ln \left[k_2 \frac{r_*}{H_{\rm QG}^{-1}}\right]
 \label{ic}
\end{equation}
where $k_1=(3^{1/2}/2^{1/3})(8\pi/3)^{1/4}\approx 2.34, k_2 = 2^{1/3}/3^{3/2}\approx 0.24$ and $r_*\equiv a_{\rm QG}x_*$.
Inverting the first equality in \eq{ic}, we can express the cosmological constant in terms of $I_c, \nu, \rho_{\rm eq}$ as:
\begin{equation}
\rho_\Lambda L_P^4 = \frac{4}{27} \ \left(\frac{3}{8\pi} \right)^{3/2} \frac{1}{\nu^6 (\rho_{\rm eq}L_P^4)^{1/2}} \ \exp \left( - 9 \pi I_c\right)
\label{four}
\end{equation}
As claimed earlier, the \textit{non-zero} value of the cosmological constant is related to the \textit{finite} value of $I_c$. The fact that even an eternal observer can only access a finite amount of information (quantified in terms of the number of modes which cross the Hubble radius) implies that the cosmological constant is non-zero; we see that $\rho_\Lambda\to0$ when $I_c\to\infty$ and vice-versa. We also see from \eq{ic} that except for a numerical factor $k_2=\mathcal{O}(1)$, the argument of the logarithm in $I_c$ is the ratio $r_*/H_{\rm QG}^{-1}$, relating the finite value of the proper size of the information horizon, $r_*$, to the finiteness of $I_c$. The region of cosmic visibility on the $a=a_{\rm QG}$ surface,  $r_*$, is finite but large (compared to $H_{\rm QG}^{-1}$) when  $\exp(3\pi I_c/2)$ is finite but large.

If $I_c$ is known  from an independent consideration, \eq{four} will determine the numerical value of the cosmological constant in terms of $(\rho_{\rm eq}, \rho_{\rm QG})$. 
To have an independent handle on $I_c$, we consider some well-established results which are fairly independent of the choice of model of quantum gravity. One such result is that \textit{the effective dimension of the quantum-corrected spacetime becomes $D=2$ close to Planck scales,} independent of the original $D$. This result was obtained, in a fairly model-independent manner (using a renormalized quantum effective metric)  in Ref. \cite{paperD}. Similar results have been established earlier by several authors (for a sample, see e.g., \cite{z1}) in a number of approaches to quantum gravity. This, in turn, implies that \cite{paperD,tp} the unit of information associated with a quantum gravitational 2-sphere of radius $L_P$ can be taken to be $I_{\rm QG}=4\pi L_P^2/L_P^2=4\pi$. With this consideration, $I_c= 4\pi$ and we obtain 
\begin{equation}
\rho_\Lambda L_P^4 = \frac{4}{27} \  \left(\frac{3}{8\pi} \right)^{3/2}\frac{1}{\nu^6 (\rho_{\rm eq}L_P^4)^{1/2}} \ \exp \left( - 36\pi^2\right)
\label{fourpi}
\end{equation} 
Given the scale $E_{\rm QG}=\nu^{-1} E_P$ at which classical geometry arises from quantum pre-geometry, the above equation determines $\rho_\Lambda$. 
At this stage, we can also reverse the argument and use the observed value of $\rho_\Lambda$ to determine the factor $\nu$. Using the result $\rho_\Lambda L_P^4 = (1.14 \pm 0.09) \times 10^{-123}$ and $\rho_{\rm eq}L_P^4 = (2.41 \pm 1.01) \times 10^{-113} $, we find that $\nu = (6.2 \pm 0.3) \times 10^{3} $ making $E_{\rm QG}$ close to the GUTs scale. These results therefore suggest that quantum gravitational effects persist for a larger range of energies than naively anticipated.

Remarkably, there is an \textit{independent} way of estimating $\nu$ by calculating the amplitude of primordial  perturbations  in terms of $\nu$, and comparing it with the observations. In the above scenario, the matter fields inherit the primordial, pre-geometric quantum fluctuations at $a=a_{\rm QG}$. 
There are two ways of estimating the resultant amplitude and spectral characteristics of the density fluctuations thus generated:  One conservative procedure is to quantize a field  in the Friedmann universe, by decomposing it into different Fourier modes,  each  labeled by the comoving wave number $k$. This will reduce the problem to that of a bunch of (time-dependent) oscillators each labeled  by $k$. A given oscillator starts in its ground state when the  quantum of (proper) energy associated with this mode, $\hbar k/a$, is equal to $E_{\rm QG}$. (This is, of course, different from choosing the Bunch-Davies vacuum for the field, as is often done in inflationary models; see e.g., \cite{wald} for a discussion). The calculation of quantum fluctuations is completely straightforward and closely parallels  the corresponding analyses for the inflationary universe. (See, for e.g. \cite{wald,tp-sriram}). The final result is given by 
\begin{equation}
 \mathcal{A} = \left[\frac{k^3P(k)}{2\pi^2}\right]^{1/2}  = \frac{c_1}{\nu} \sqrt\frac{4}{3 \pi} \left[\frac{3 w^{1/2} (6 w + 5)}{4 (3w + 5)^2}\right]^{1/2} = \frac{0.19 c_1} {\nu}
 \label{eqn6}
\end{equation} 
for $w=1/3$, where $c_1$ is a numerical factor of order unity whose exact value can be determined by more detailed analysis.\footnote{This can be obtained, for example, from eq.(16) of Ref. \cite{wald}, taking care of the fact that $l_p^2$ in Ref. \cite{wald} is $(8\pi/3)L_P^2$ and $l_0=\nu L_P$.} Using the value of $\nu$ determined from \eq{fourpi}, we find that 
 $\mathcal{A}_{\rm theory} = 3.05 c_1 \times 10^{-5}$  which has to be compared with the observed value $\mathcal{A}_{\rm obs} \approx 4.69  \times 10^{-5}$. We see that the results are remarkably consistent with $c_1=1.54=\mathcal{O}(1)$.
 
 A more speculative -- and exciting -- possibility is to generate the perturbations directly from the quantum pre-geometric phase \cite{lee}. This uses the fact that if the pre-geometric phase obeys holographic equipartition \cite{tp}, it can be modeled as a thermal system with energy $E\propto A T_c$ where $T_c\approx E_{\rm QG}=E_{\rm Pl}/\mu$ is the critical temperature at which the quantum to classical transition occurs and $A\propto R^2$ is the area of the boundary. Such a system has a specific heat  $C\propto A \propto R^2$ leading to energy fluctuations $\sigma_E^2 = CT^2 \propto A\propto R^2$.
This, in turn, leads to perturbations in the energy density $\delta \rho  = \delta E/V $ such that $\sigma^2_\rho = \sigma^2_E/V^2\propto\sigma^2_E/R^6$. It can be shown that this will lead (see Ref. \cite{lee} for details; for similar ideas, see e.g., Ref. \cite{related}) to a scale invariant  spectrum with $\mathcal{A} \approx T_c/E_{\rm Pl}\approx\nu^{-1}$. We see that the observed result for $\mathcal{A}$ is again obtained when  $\nu \approx \mathcal{O}(1) \times 10^4$. In this analysis, we thus have a clear identification of a transition from the pre-geometric phase to geometric phase occurring at the energy scale $\nu^{-1} E_{\rm Pl}$, with consistent results.

We will now elaborate on some of the ingredients which have gone into the results,
which emphasize the underlying logical structure of the framework.

 We consider a universe which 
 makes a transition from a quantum, pre-geometric phase to the classical geometric description at $a=a_{\rm QG}$ when the characteristic energy scale is $E_{QG}\equiv \nu^{-1} E_{\rm Pl}$. Our aim is to connect the quantum and classical phases using the concept of information accessible to an eternal observer.  
To explore such a paradigm based on cosmic information, we first have to define it. We \textit{define} the relevant quantity, $I_c$,  using the result in \eq{defN}, as the  number of length scales which enter the Hubble radius during the history of the universe. 

Demanding that $I_c$ should be finite \textit{requires} the Hubble radius to have a maximum at some $a=a_\Lambda$ so that $I_c\equiv N(a_\Lambda,a_{QG})$ is finite. We should have $(\rho+3p)=0$ at $a=a_\Lambda$, followed by a phase of accelerated expansion when $(\rho+3p)<0$.  If we do not introduce any exotic, untested physics, then the  simplest model exhibiting $(\rho+3p)<0$ at late times is the one with a non-zero cosmological constant.
So we are led to a model with matter, radiation and a cosmological constant and no other exotic forms of matter, either in the early phase or at the late stages of evolution.
We then relate, purely algebraically, (i.e., without any additional assumptions) the $I_c$ 
to $\rho_\Lambda$ [see \eq{four}] with $\rho_\Lambda\to0$ when $I_c\to\infty$ and vice-versa. This connects the information content to the cosmological constant. The model is also capable of generating scale invariant primordial perturbations with an amplitude  $\mathcal{A}\approx \nu^{-1}$. This has been worked out in  two different but viable scenarios, one fairly conservative \cite{wald} and the other more speculative \cite{lee}.
The choice of $\nu \approx 10^{4}$ leads to the correct value for \textit{both} $\mathcal{A}$ and $\rho_\Lambda L_P^4$. 

It is known from standard inflationary calculations that $\mathcal{A} \sim E_{\rm inf}/E_{\rm Pl}$ where $E_{\rm inf}$ is the energy scale of inflation. It is therefore expected that $\nu^{-1} \approx 10^{-4}$ gives the correct amplitude for the perturbations. But the key new discovery is that the \textit{same} value of $\nu$ leads to the precise, observed value of the cosmological constant.  
That is, we determine \textit{two} quantities $\mathcal{A}$ and $\rho_\Lambda L_P^4$ ---  neither of which can be determined from first principles in conventional cosmology --- from a \textit{single} parameter $\nu$. There is no a priori reason why a specific value for $\nu$ should lead to the correct, observed values for both $\mathcal{A}$ and $\rho_\Lambda L_P^4$. \textit{This is the strongest argument in favour of this scenario.}

Another way of expressing this key result is to note that
\begin{equation}
 I_c = -\frac{2}{3\pi}\ln\left[\frac{k_1 (\rho_\Lambda^2\rho_{\rm eq})^{1/12}}{E_{\rm QG}}\right] = 4\pi [1+\mathcal{O}(10^{-3})]
 \label{eqn7}
\end{equation} 
when $\nu$ has the value \textit{determined by the observed amplitude} of the primordial  spectrum. The fact that the specific combination of parameters defining $I_c$ has a simple value equal to $4\pi$ (to the accuracy of one part in a thousand!) cries out for an explanation. This result is naturally obtained by identifying $I_c$ with the information accessible to the eternal observer and $4\pi$ with the quantum gravitational unit of information. (Note that we do \textit{not} ``fix'' $\nu$ and $E_{\rm QG} = \nu^{-1} E_{\rm Pl}$ such that \eq{eqn7} holds; instead $\nu$ can be determined from \eq{eqn6}.)

In the standard approach to theories of gravity interacting with matter fields, varying the metric tensor leads to the gravitational field equations in the form $G_{ab} = \kappa T_{ab}$ (where $G_{ab}$ is proportional to the Einstein tensor in general relativity, but could be a more complicated tensor in a general theory like, e.g., the \LL\ models).
This  field equation is clearly \textit{not} invariant under the addition of a constant to the matter Lagrangian. This is equivalent to the introduction of a cosmological constant (if it was not present originally), or a change in its numerical value. Therefore, in such an approach, any physical principle to determine the value of the cosmological constant is dubious.
The cosmological constant problem can thus be solved \textit{only if} the gravitational field \textit{equations} are made invariant under the addition of a constant to the matter Lagrangian, but their \textit{solutions} permit an inclusion of the cosmological constant.
This is accomplished naturally in the emergent gravity paradigm, in which the field equations of gravity are invariant under the addition of a constant to the matter Lagrangian. It can be shown that the cosmological constant arises as an integration constant in the solutions. It is possible to reformulate the  GR (and in fact, also its extension to the \LL\ models), using the emergent gravity paradigm.
A new physical principle is therefore required to fix the numerical value of the integration constant, i.e. the cosmological constant. This is exactly what is achieved in this paper and furthermore, connects the value of the cosmological constant to cosmic information and the amplitude of primordial perturbations. This issue has been addressed extensively in several previous papers on the emergent gravity paradigm;  see for e.g., Ref. \cite{hptpreview,tp}.

Our approach does not invoke inflation in the standard manner with inflaton fields.
Conventional cosmology requires the  inflationary paradigm only to produce a scale invariant primordial spectrum \cite{comment4}.  The other ``problems'' which inflation is supposed to ``solve'' cannot be considered sufficient motivation for inflation. (For example, 
the quantum correlations in the pre-geometric phase can solve the conventional horizon problem in this approach.) In fact, the generation of the primordial spectrum in the models  mentioned above \cite{lee,wald} uses a single parameter to predict the spectrum --- which is  conceptually superior to the plethora of models with various fine-tuned potentials $V(\phi)$ for the inflaton fields. The details of these (and similar) models need to be worked out further (e.g., as regards the tensor-to-scalar ratio, taking QG effects into account \cite{tpgw}) to provide a more complete picture; but these initial  results are extremely promising. 

This work makes three distinct improvements on our earlier work \cite{hptpreview, hptpearlier} linking CosmIn and the cosmological constant: (i) We do not require an inflationary model or its energy scale. Instead, we obtain the results from a model involving minimal assumptions about the quantum to classical transition of the universe \cite{comment5}. (ii) We show that \textit{both} the cosmological constant \textit{and} the amplitude of the perturbation spectrum can arise naturally in such a model. (iii) We provide a quantum gravitational motivation for using the area ($4\pi$) of a unit 2-sphere, rather than the area of a unit $D$-sphere, as the quantum of information based on Ref.\cite{paperD} and  others \cite{z1}. 

The results here bring to center-stage the notion of spacetime information and its role in gravitational dynamics, already seen in several other contexts \cite{tp}. It also strengthens the viewpoint, suggested in Refs. \cite{tp,lwf}, that the universe should not be treated as a particular solution to the gravitational field equations but instead, be approached as a special dynamical system.

Finally we emphasise that all our results follow from one single \textit{definition} (of $N(a_2,a_1)$ in \eq{defN}) and the postulate 
$N(a_\Lambda, a_{\rm QG}) = 4\pi$
where $N(a_\Lambda, a_{\rm QG})$,  is the total (maximum) number of modes which enter the Hubble radius from the time the universe made a transition to classicality ($a_{\rm QG}$) up to the epoch $a_\Lambda$, until when the modes continue to enter the Hubble radius.  
Given this single assumption and the fact that $4\pi$ is finite, it follows that $N(a_\Lambda, a_{\rm QG})$  as well as $a_\Lambda$ have to be finite. This, in turn, \textit{requires} a turn around in the Hubble radius and leads to a late time acceleration phase.  
Computing $N(a_\Lambda, a_{\rm QG})$ for a universe with radiation, matter and the cosmological constant, and using $N(a_\Lambda, a_{\rm QG}) = 4\pi$, we obtain \eq{fourpi} of the paper. Previous work cited \cite{wald,lee} leads to \eq{eqn6} of the paper. We find that we can satisfy \eq{fourpi} and \eq{eqn6} with a single value of $\nu$, which is the main result of the paper.
So, given a single assumption (viz., $N(a_\Lambda, a_{\rm QG})=4\pi$), we can derive all the key conclusions of the paper. Further, the validity of \eq{fourpi} and \eq{eqn6} from observations tells us that this assumption is indeed true (to the accuracy of one part in a thousand, as mentioned in \eq{eqn7}). 
Obviously, we need to understand how the postulate $N(a_\Lambda, a_{\rm QG})=4\pi$ using the definition of $N(a_\Lambda, a_{\rm QG})$, based on counting the modes by $d^3x\ d^3k/(2\pi)^3$,  relates to other notions of information used in quantum gravity. It appears that cosmology requires a specific approach to quantum information. We hope future work on QG and emergent gravity will throw light on this.

\textit{Acknowledgements:} The research work of TP is partially supported by the J.C. Bose research grant of DST, India. The research of HP is supported by the Tomalla Foundation.



\begin{thebibliography}{000}
\bibitem{comment1} 
We do \textit{not} assume that $\nu$ is of order unity, thereby allowing the possibility that quantum gravitational effects can have a long tail.  The transition could be sudden (e.g. like a phase transition) or gradual, and in the latter case $\nu^{-1} E_{\rm Pl}$ is equivalent to the \textit{effective} scale at which the transition can be approximated as a sudden occurrence.
\bibitem{comment3}
This quantity  $x_\infty(0)\approx x_\infty(a_{\rm QG})$
can be expressed in terms of an elliptic integral. 
For $(\rho_\Lambda/\rho_{eq})^{1/4}\approx 2.62 \times 10^{-3} $, we find that $a_{eq} H_\Lambda x_* = 9.99 \times 10^{-4}$ which is the maximum comoving distance we can ever probe. We have already probed a fraction 0.74 of this today.
\bibitem{comment5} The preceding analysis will, of course, go through even if we assume that there was a inflationary phase just prior to $a=a_{\rm QG}$, with the Hubble constant $H_{\rm QG}$. But conceptually, a more satisfying picture --- as well as a strong motivation for $I_c=4\pi$ --- arises from assuming the universe made a transition at the energy scale $E_{\rm QG}$. Inflation need not be invoked in this scenario.
\end{thebibliography}
\end{document}